\documentclass[aps, superscriptaddress,groupedaddress]{revtex4}  

\usepackage{amsmath, amsthm, amssymb}
\usepackage{graphicx,epsfig}
\usepackage{color}
\usepackage{natbib}
\usepackage{setspace}
\usepackage{fullpage}
\usepackage{sidecap}

\begin{document}
\begin{spacing}{2}

\title{Effect of oxidation on the Mechanical Properties of Liquid Gallium and Eutectic Gallium-Indium}

\affiliation{Department of Physics and James Franck Institute, University of Chicago, Chicago, IL 60637, USA}
\affiliation{Physics of Fluids Group, Department of Science and Technology, University of Twente, 7500 AE Enschede, The Netherlands}
\affiliation{School of Natural Sciences, University of California, Merced, CA 95343, USA}

\author{Qin Xu} \affiliation{Department of Physics and James Franck Institute, University of Chicago, Chicago, IL 60637, USA}
\author{Nikolai Oudalov} \affiliation{Physics of Fluids Group, Department of Science and Technology, University of Twente, 7500 AE Enschede, The Netherlands}
\author{Qiti Guo} \affiliation{Department of Physics and James Franck Institute, University of Chicago, Chicago, IL 60637, USA}
\author{Heinrich M. Jaeger} \affiliation{Department of Physics and James Franck Institute, University of Chicago, Chicago, IL 60637, USA}
\author{Eric Brown} \affiliation{Department of Physics and James Franck Institute, University of Chicago, Chicago, IL 60637, USA}
\affiliation{School of Natural Sciences, University of California, Merced, CA 95343, USA}

\date{\today}

\begin{abstract}
Liquid metals exhibit remarkable mechanical properties, in particular large surface tension and low viscosity. However, these properties are greatly affected by oxidation when exposed to air.  We measure the viscosity, surface tension, and contact angle of gallium (Ga) and a eutectic gallium-indium alloy (eGaIn) while controlling such oxidation by surrounding the metal with an acid bath of variable concentration.  Rheometry measurements reveal a yield stress directly attributable to an oxide skin that obscures the intrinsic behavior of the liquid metals.  We demonstrate how the intrinsic viscosity can be obtained with precision through a scaling technique that collapses low- and high-Reynolds number data.  Measuring surface tension with a pendant drop method, we show that the oxide skin generates a surface stress that mimics surface tension and develop a simple model to relate this to the yield stress obtained from rheometry.  We find that yield stress, surface tension, and contact angle all transition from solid-like to liquid behavior at the same critical acid concentration, thereby quantitatively confirming that the wettability of these liquid metals is due to the oxide skin.
\end{abstract}
\maketitle

\section{Introduction}
Liquid metals combine the largest surface tension of any fluid with high density and conductivity$^{1-3}$.  Several of these metals, in particular liquid gallium and its alloys, are desired also for their low melting points and their wetting properties and have been used for thin film coatings$^{4,5}$, as moldable electronic interconnects$^6$, or as penetrating contrast agents for the visualization of grain boundaries or micro-cracks$^{7-9}$. However, the dynamical properties of liquid metals are significantly more complex than those of simple Newtonian liquids. At small applied stresses liquid gallium and its alloys show a solid-like elastic response$^{10,11}$, and high-temperature molten metals under steady state shear have been found to exhibit non-Newtonian behavior$^{12}$. It has been suggested$^{10,11}$ that the solid-like properties are due to oxidation when the metal is exposed to air.  It also has long been known$^{10}$ that gallium, which is highly wetting when oxidized, becomes non-wetting on glass when oxidation is prevented. A study with quantitative control of the oxidation conditions has not been performed so far.  In particular, it has remained unclear if the desirable intrinsic properties of the pure liquid such as viscosity can be obtained along with wetting which is associated with oxidation.

To address this issue, we present a systematic investigation of two liquid metals, gallium (Ga) and a eutectic gallium-indium alloy (eGaIn). This study combines detailed measurements of the dynamic response, obtained by rheometry, with measurements of surface tension and contact angle.  We control oxidation by surrounding the liquid metal in a bath of variable acid concentration. The acid bath not only prevents oxidation by mitigating contact with the air, but initiates a reduction process that can remove an oxide surface layer. By varying the acid concentration, we can precisely and reproducibly tune the balance between oxidation and reduction and map out the transition from solid-like to liquid behavior. This allows us to quantitatively  analyze the role played by the oxide layer in determining the overall mechanical properties.

The remainder of the paper is organized as follows.   In Sec.II. we describe the materials and rheometer setup and measurement procedures,  In Sec. III. A.1 we present measurements of shear stress over time to demonstrate how oxidation and reduction at different acid concentrations result in time-dependence of shear stress at a fixed shear rate. In Sec. III. A. 2\&3 we present steady state measurements of shear stress vs. shear rate, then introduce a new scaling technique to obtain the intrinsic liquid viscosity  in the presence of non-Newtonian effects from rheometer data at higher Reynolds numbers.  In Sec. III. B we show measurements of the effective surface tension using the pendant drop method and develop a model to identify the contribution of yield stress to drop size.  Contact angle data are provided in Sec. III. C. This section also compares the yield stress, surface tension, and contact angle measurements as a function of acid concentration, demonstrating that the transition from solid-like to liquid behavior in all cases occurs at the same critical acid concentration. Sec. IV contains a summary and conclusions.

\section{Experimental}
\subsection{Materials}
We purchased Ga (purity 99.99$\%$) and Indium (In, purity 99.995$\%$) for our experiment from Sigma-Aldrich, with the melting points around 30$^{\circ}$C and 156$^{\circ}$C respectively.   We produced eGaIn by mingling 77\% Ga with 23\% In. The mixture was kept at a temperature of 200$^oC$ in a quartz tube for about two hours, while argon gas was cycled through the tube to inhibit oxidation. The liquid form of eGaIn (melting point $\sim$ 15$^oC$) was obtained after cooling down the mixture to room temperature.  All measurements of mechanical properties were performed at 50$^{\circ}$C in the liquid state.  At this temperature, Ga has a density of $\rho = 6.095g/ml$ and nominal viscosity of ${\eta_l = 1.90\times10^{-3}}mN\cdot s$[1], and eGaIn has a density of $\rho = 6.25g/ml$ and nominal viscosity of ${\eta_l = 1.86 \times10^{-3}}mN\cdot s$[10]. For the acid bath, we used hydrochloric acid (HCl) of varying concentration. 

\subsection{Methods}

\begin{figure}
\begin{center}
\includegraphics[width=155mm]{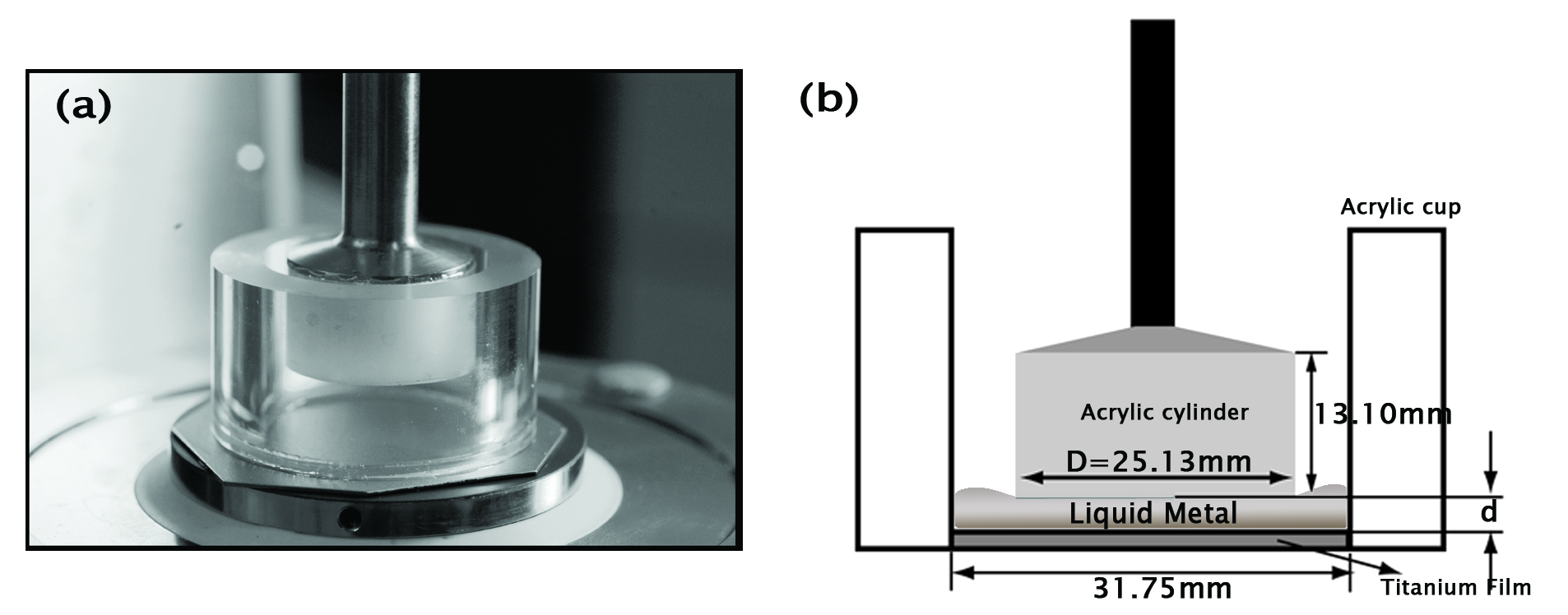}
\end{center}
\caption{(a) A picture of the rheometer set up. (b). A schematic illustration with dimensions.}
\end{figure}

To measure the viscosity of liquid metals, we used an Anton Paar rheometer with a parallel-plate geometry. Fig.1 shows the experimental setup, with plate diameter $D= 25.13$ mm and gap size $d$ which was varied from 0.2 to 2 mm for different measurements. The rheometer measures the torque $T$ exerted to rotate the top plate and the angular rotation rate $\omega$.  These are used to calculate the average shear stress at the plate edge $\tau=16T/\pi D^3$ and average shear rate at the edge $\dot\gamma=D\omega/2d$.  The stress resolution is 0.03 Pa.  The effective dynamic viscosity is $\eta_{eff}=\tau/\dot{\gamma}$ in the steady state.  To contain the acid bath, a cup was constructed consisting of a clear cast acrylic cylinder with a 31.75 mm inside diameter glued to a titanium plate of 0.51mm thickness that formed the bottom.  The cup also helps contain the liquid metal, which due to their high surface tension with the tool surfaces do not stick.  For this reason, we fill the liquid metal to a height larger than the gap size $d$ so there is no space for the liquid to spill outward.  Additionally, a 13.10 mm thick acrylic cylinder is attached to the tool to protect it from both the acid and gallium.  At shear rates above $\sim 10^4s^{-1}$, the surface shape of the liquid starts to change from a variety of instabilities, so we only report measurements at lower shear rates where the geometry of the liquid region remains fixed. 

\section{Results}
\subsection{Viscosity}
\subsubsection{Oxidation and reduction on the surface}
\label{sec:oxidation}

\begin{figure}[h]
\begin{center}
\includegraphics[width=85mm]{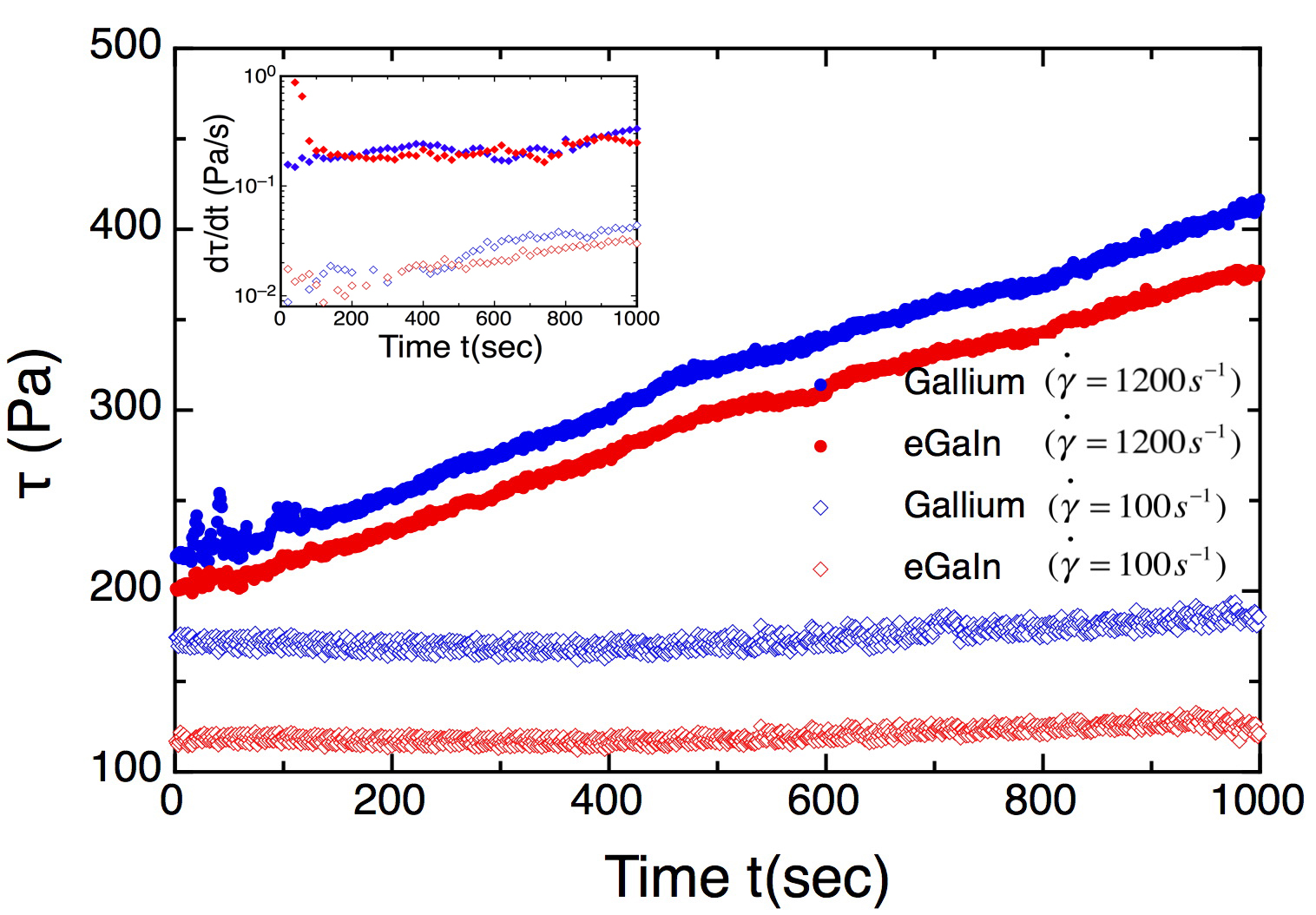}
\end{center}
\caption{Oxidation of liquid metals exposed to air in rheometer measurements. The time evolution of shear stress $\tau$ for both Ga (blue) and eGaIn (red) at $\dot{\gamma}=100s^{-1}$ (open symbols) and $\dot{\gamma}=1200s^{-1}$ (closed symbols).  The increase of $\tau$ with time at $\dot{\gamma}=1200s^{-1}$ implies ongoing solidification by oxidation at the surface. Inset: plot of $d\tau/dt$ vs. time.}
\end{figure}

We first demonstrate the effect of the oxidation process on the rheology of liquid metals.  In these experiments we measured the time dependence of shear stress $\tau$ at constant shear rate $\dot{\gamma}$ as oxidation and reduction processes may be taking place, under different initial conditions, both in air and in the acid bath, and for both Ga and eGaIn. 

Initially, 2ml of oxidized liquid metal was directly placed on the bottom plate of the rheometer with a pipette and exposed to the air, and then rested for around 30s to allow the shape of the surface to fully relax.  Such an oxidized droplet appears dirty and wrinkled on the surface, while the bulk appears pure if we slice open the skin. Care was taken to avoid any agitation or shear that could break the layer before the experiment.  We then lowered the tool to a gap size $d=2$ mm and started to shear the sample with a constant shear rate. Fig.~2 shows the time-evolution of shear stress for experiments at different shear rates. The existence of the solid oxide layer is seen to initially cause a large stress $\tau\sim 10^2$ Pa, which is two orders of magnitude larger than the expected viscous contribution of the bulk$^{1,10}$ for $\dot\gamma=100$ s$^{-1}$. At this shear rate, $\tau$ displays very slow increase with time, indicating that the sample is almost in equilibrium.  However, at $\dot{\gamma}=1200s^{-1}$, $\tau$ increases linearly with time. It can be seen that the oxide skin is broken at the higher shear rate and the surface layer turns over so that the bulk material becomes exposed to the air and has a chance to react with the oxygen, creating more solid oxidized material to resist the rotation of the rheometer plate. This result can be explained by continuing solidification by oxidation happening on the surface under shear.

To demonstrate the role of the acid bath in reducing the oxidized surface layer, we performed a similar experiment in acid baths of varying concentrations  of HCl.  In these experiments, 2ml of previously oxidized Ga (or eGaIn) was placed with a pipette into the acrylic cup (Fig.1) filled with HCl solution.  A constant shear rate was imposed at $t=0$ immediately after the liquid metal was placed in the acid. Fig.3 (a) shows the time-evolution of the shear stress for $\dot{\gamma}=60s^{-1}$ for several different acid baths with acid concentrations $c_{\mbox {\scriptsize HCl}}$.  In contrast with the measurements with exposure to air, now the shear stress starts at low values, decreases with time, and eventually reaches a steady state.  This suggests the acid bath not only inhibits further oxidation but actually reduces the pre-existing oxide layer, and eventually an equilibrium value is reached depending on concentration where oxidation and reduction processes are in balance.  The equilibrium stress is seen to decrease with $c_{\mbox {\scriptsize HCl}}$ and level off for $c_{\mbox {\scriptsize HCl}} \stackrel{>}{_\sim} 0.5$ M.  Equilibrated liquid Ga in acid baths with higher $c_{\mbox {\scriptsize HCl}}$ appears to be very shiny, comparable to a mirror finish, which is expected for the pure liquid.  The remaining stress in equilibrium can be attributed to hydrodynamic stresses in the bulk. This suggests that a pure liquid state of Ga can be reached by vigorous washing of the liquid metal in HCl at a concentration $c_{\mbox {\scriptsize HCl}} \stackrel{>}{_\sim} 0.5$ M.

\begin{figure}
\begin{center}
\includegraphics[width=155mm]{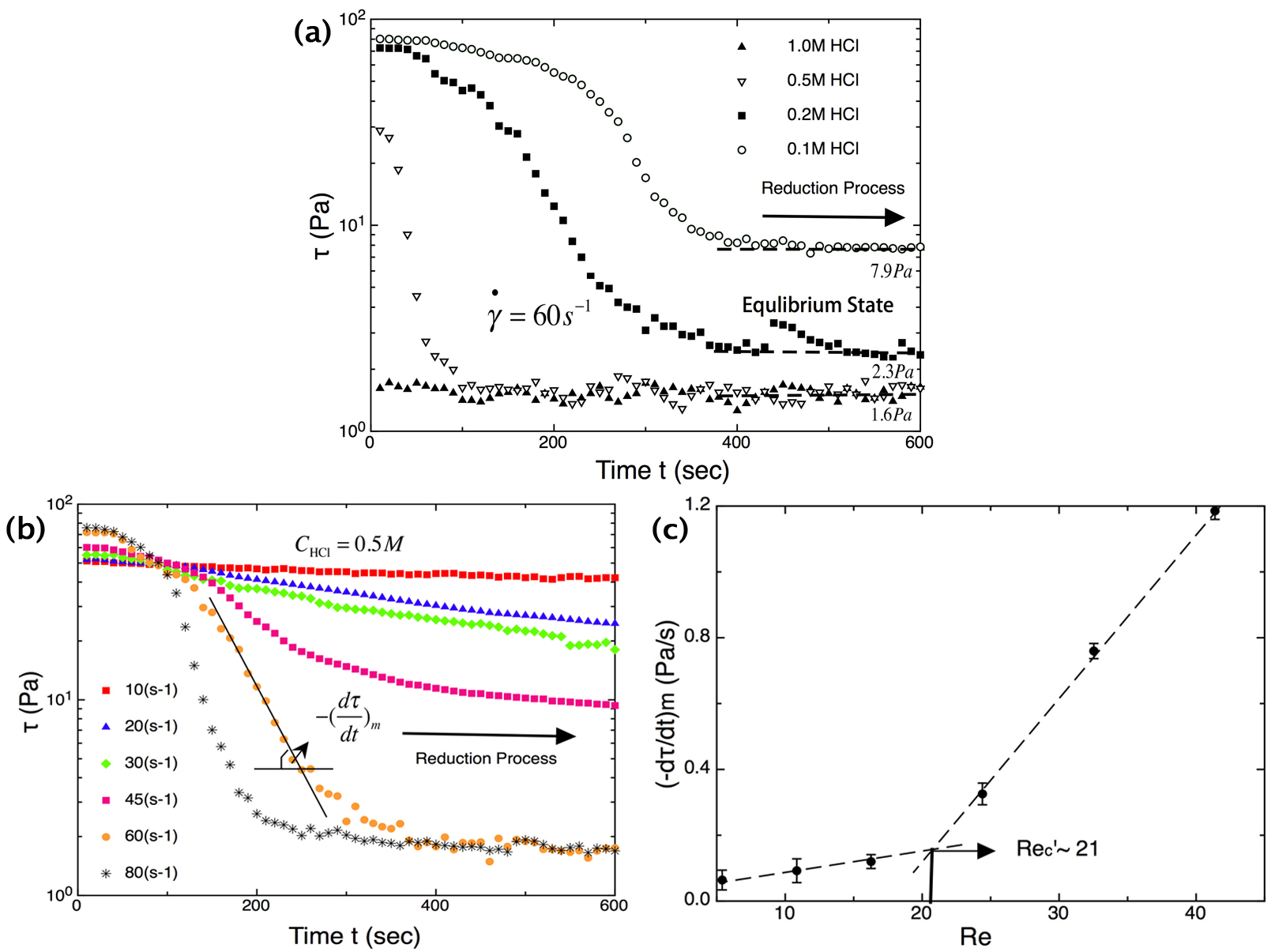}
\end{center}
\caption{Reduction of the oxide layer for gallium in rheometer measurements with an acid bath.  (a) Time evolution of shear stress $\tau$ at $\dot\gamma=60$ s$^{-1}$ for different acid concentrations.  (b) Time evolution of shear stress $\tau$ at different shear rates for $c_{\mbox {\scriptsize HCl}} = 0.5$ M.  (c).  Maximum rate of change of stress $-(\frac{d\tau}{dt})_m$ from panel (b) plotted against Reynolds number. Here, dashed lines represent two different linear scaling regimes, whose intersections defines a transitional Reynolds number, $Re_c^\prime\sim21\pm 2$, above which the reduction process speeds up dramatically. 
}
\end{figure}

The time-evolution of stress for additional measurements at different shear rates with $c_{\mbox {\scriptsize HCl}} = 0.5$ M are shown in Fig.~3 (b).  The rate of stress decrease is seen to increase with $\dot{\gamma}$.  This conforms what we observed for the oxidation measurements: faster shear increases the exposure of the bulk liquid metal to the surrounding environment and accelerates the corresponding reactions. To quantify the relationship between reduction rate and shear speed, we calculate $(-d\tau/dt)_m$ as the slope of the steepest portion along each curve, and plot this slope vs Reynolds number $Re=\rho d^2 \dot{\gamma}/\eta_l$ in Fig.3 (c).  There appear to be two linear scaling regimes, with a steeper slope at higher $Re$.  We fit the two regimes  by straight lines, whose intersection defines a transitional Reynolds number ${Re_c}^\prime\sim 21\pm 2$ above which the rate of stress increase by oxidation increases rapidly. We attempted a similar measurement for the shear rate dependence of the oxidation in air (Fig. 2).  However, the oxidation in air happens much slower than reduction in the acid bath (by a factor of $\sim 10^2$) so that we could not resolve a similar transition for oxidation. In Sec. III. A. 3, we will show that inertial flow is seen for  $Re>Re^{\prime}_c$, suggesting the enhanced reaction rate is the result of enhanced mixing from vortical flows at high $Re$.

\subsubsection{Viscosity curves}

\begin{figure*}
\begin{center}
\includegraphics[width=170mm]{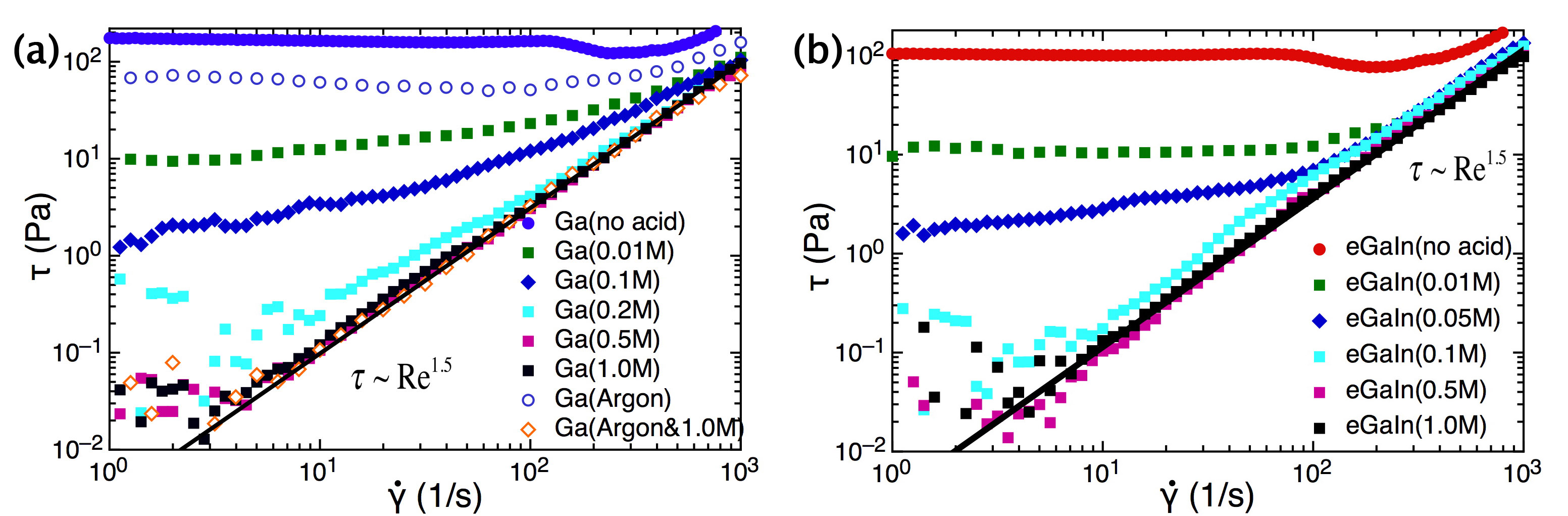}
\end{center}
\caption{Shear stress $\tau$ vs shear rate $\dot{\gamma}$ in the steady state for (a) Ga and (b) eGaIn.  HCl bath concentrations are shown in the keys.  Solid circles: liquid metals are exposed to air without an acid bath.  Open symbols:  Gallium in an Argon atmosphere without an acid bath, in which the gallium was first brought to equilibrium either in air (open blue circles) or a 1 M HCl bath (open yellow diamonds).  Solid line: $\tau \propto \dot\gamma^{3/2}$ corresponding to intermediate Reynolds number flow.
}
\end{figure*}

We next characterize the relation between shear stress $\tau$ and shear rate $\dot\gamma$ for Ga and eGaIn at different atmospheric and bath conditions in the steady state.  To achieve chemical equilibrium, the samples were pre-sheared before each measurement at $\dot\gamma=60s^{-1}$ for  around 600s according to Fig.~3.  The steady state was confirmed by slowly ramping the shear rate first up then down, and checking that there was no hysteresis in the measured stress, except when the liquid metal is exposed to air where the oxidation is not reversible by shear. Figure 4 shows $\tau$ vs $\dot\gamma$ at for different acid bath concentrations $c_{\mbox{\scriptsize HCl}}$ at a gap size $d= 2$ mm. Data are shown for Ga in panel (a) and for eGaIn in panel (b).  When exposed to air, the liquid metals are seen to exhibit a yield stress, i.e, the shear stress goes to a non-zero value in the limit of zero shear rate.  This indicates solid behavior from the oxidized layer, since at stresses below the yield stress, the material does not flow.   This value is consistent with oscillatory rheometer measurements$^{11}$.  

Adding the acid bath and increasing the HCL concentration reduces this yield stress.   The yield stress plateaus at a minimum value for HCl concentrations greater than about 0.5M for both metals.  Along with the data of Figs. 2 and 3, this confirms that an HCl concentration greater than about 0.5M is enough to minimize the contribution of the oxidation to the shear stress. 
 
Another major feature of Fig.~4 is that the stresses collapse onto a single curve at high shear rates.  This curve extends down to lower shear rates for higher acid concentrations where the yield stress is lower.  This behavior suggests the total stress comes from the sum of two components: the yield stress which is dependent on the acid concentration, and a hydrodynamic contribution dependent on the shear rate.  The latter component scales as $\tau \propto \dot\gamma^{3/2}$, which differs from the laminar regime of a Newtonian fluid.  We will show in the next section that this behavior is the hydrodynamic shear stress for intermediate Reynolds number flows where inertial effects play a role.

So far we have used the acid bath to prevent oxidation, but an alternative method would be to use an inert atmosphere. However, Ref.1 points out that argon gas environment can not protect the the surface of gallium from oxidation. To test this statement, we sealed a plastic bag around the rheometer and filled it with argon gas to replace the air. Argon gas was pumped into a larger cylinder surrounding the sample, and air was allowed to escape through a hole at the top.  Liquid Ga samples were first washed in an acid bath to remove all oxidation, then placed within the rheometer in the argon atmosphere.  A plot of $\tau$ vs $\dot\gamma$ for this sample is shown as open orange diamonds in Fig.~4(a).  This sample behaves similar to those submerged in high-concentration acid baths.  This test lasted well over 1hr, over which time oxidation would have occurred to produce a significant yield stress if this experiment had been done in air.  Therefore our results confirms that an inert atmosphere can prevent oxidation just as well as an acid bath, which is in disagreement with claims in Ref.1. If instead a sample of liquid Ga that is already oxidized from exposure to air is placed in the rheometer with the argon atmosphere, the plot of $\tau$ vs $\dot\gamma$ shown as open blue circles in Fig.~4 (a) exhibits a large yield stress, similar to the case of a sample measured in air.  This illustrates that, while an inert atmosphere can prevent further oxidation, it cannot reduce a pre-existing oxidation layer like an acid bath can.
 
\subsubsection{Reynolds number scaling}
\label{sec:inertialscaling}

\begin{figure}
\begin{center}
\includegraphics[width=155mm]{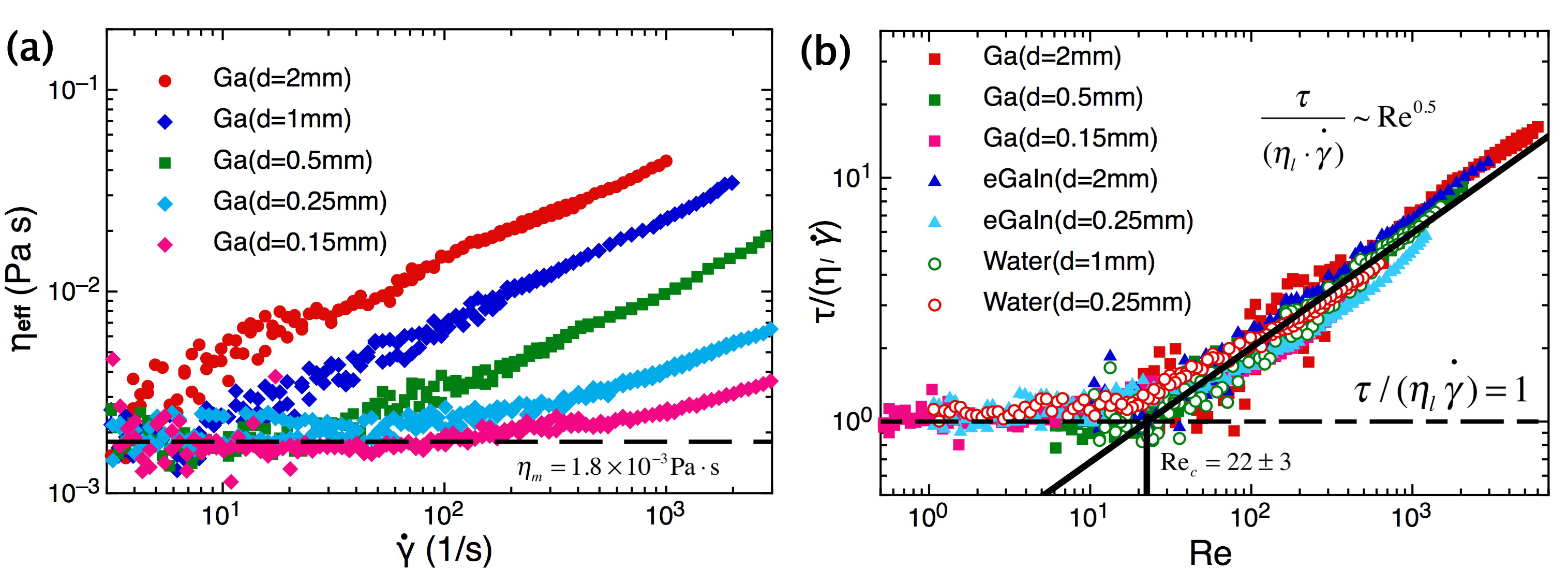}
\end{center}
\caption{Reynolds number collapse of stress measurements. (a). Viscosity $\eta_{eff}= \tau/\dot\gamma$ vs. shear rate $\dot\gamma$ for liquid Ga in a 1.0M HCl bath where there is no oxide layer. The gap size $d$ is shown in the key (b). The rescaled shear stress $\tau/(\eta\dot{\gamma})$ plotted against $Re=\rho_l\dot\gamma d^2/\eta_l$ for Ga, eGaIn, and water.  The data collapse in $Re$ identifies two scaling regimes: the laminar regime at low $Re$ where $\tau \propto \dot\gamma$ and an intermediate $Re$ regime where $\tau \propto \dot\gamma^{3/2}$.  The intersection of these two regimes identifies a critical Reynolds number $Re_c=22\pm3$ for our measuring geometry.}
\end{figure}

In this section, we demonstrate that rescaling $\tau(\dot\gamma)$ using the Reynolds number can be used to precisely measure the low intrinsic viscosities of liquid metals.  The effective viscosity of a fluid is defined as  $\eta_{eff} = \tau/\dot{\gamma}$.  When there is an oxide layer, $\eta_{eff}$ is dominated by the yield stress at low shear rates.  To avoid this effect, we concentrate on measurements in a 1.0M HCl bath, and plot $\eta_{eff}$ vs $\dot\gamma$ for Ga in Fig.~5 (a) for different gap sizes $d$.  In the limit of low shear rates, for the smallest gap sizes, $\eta_{eff}$ is constant, corresponding to the intrinsic bulk viscosity $\eta_l\sim(1.82\pm0.21)\times 10^{-3}$ Pa$\cdot$s.  

This Newtonian regime is difficult to resolve in a rheometer for low viscosity fluids because the stress resolution limit leads to relatively large uncertainties at low shear rates,  and many rheometers would not be able to resolve this regime.  To our knowledge this is only the second direct rheometer measurement of the intrinsic viscosity of a liquid metal$^{12}$.  The Newtonian regime may also be unresolvable in the presence of non-Newtonian effects such as a yield stress; we have already omitted our data with a yield stress from Fig.~5, but a similar scaling technique could also be applied to such data.  In fact, we were only able to resolve the Newtonian regime at small gap sizes.  At higher shear rates or larger gap sizes, the Reynolds number is larger so inertial effects increase energy dissipation and $\eta_{eff}$ increases with $\dot\gamma$ as seen in Fig.~5 (a).  

To confirm that the observed increase in $\eta_{eff}$ with shear rate is an inertial effect, we rescale the data in Fig.~5 (a) non-dimensionally by dividing the stress by the viscous  contribution and $\tau/\eta_l\dot\gamma$ and converting the shear rate to the Reynolds number $Re={\rho d^2\dot{\gamma}}/{\eta_l}$ and replot it in Fig.~5 (b).  It is now seen that the data for different gap sizes collapse onto each other, confirming that this dimensionless scaling is appropriate to describe the fluid flow. Particularly, at small gap sizes (light blue and pink points in Fig. 5), the plots extend along the constant viscosity region with a large range of shear rate, which thus can be used to determine viscous regime. By contrast, data at large gap sizes (red, dark blue and green points) is at a higher Reynolds number and thus larger stresses and a better stress resolution.  Since this rescaling depends only on Newtonian fluid properties, it should apply to any Newtonian fluid in the same flow geometry. To test this, we also show measurements for water and eGaIn in the same flow geometry in Fig.~5 (b).  The data for each liquid collapses onto the same curve, confirming that this rescaling applies to all Newtonian fluids in the same flow geometry.

Since rheometer measurements have a better resolution in the high Reynolds number regime, we can take advantage of the universal scaling collapse to infer the intrinsic liquid viscosity from this regime by calibrating with a liquid of known viscosity.  In the high Reynolds number regime ($30 < Re < 6000$), the dimensionless stress is found to follow the empirical relation $\tau/(\eta_l\dot\gamma) \propto Re^{1/2}$ for a fit parameter $a$, as shown in Fig.~5 (b), which is similar to other flows in a similar range of Re as a result of increasing inertial forces$^{13}$.  Combining this scaling regime with the viscous scaling $\tau = \eta_l\dot\gamma$ in the limit of low $Re$ gives a relationship where the two scaling laws meet at a critical Reynolds number $Re_c$ defined by

\begin{equation}
\frac{\tau}{\eta_l\dot\gamma}=\left(\frac{Re}{Re_c}\right)^{1/2}.
\end{equation}

\noindent  Rearranging this gives a relationship gives an expression for $\tau(\dot\gamma)$

\begin{equation}
\tau=\left(\frac{\rho_l \eta_l \dot\gamma^3 d^2}{Re_c}\right)^{1/2}.
\label{eqn:stressfit}
\end{equation}

\noindent Fitting this equation to $\tau(\dot\gamma)$ gives the ratio between $\eta_l/Re_c$.  This process can be applied to a Newtonian fluid in the same geometry with known viscosity $\eta_l$ to obtain $Re_c$, so that $\eta_l$ can be determined for other fluids.  

Since the empirical inertial scaling is only valid in a limited range of $Re$, it is necessary that these fits are done in the same range of $Re$ for different fluids.  In our case, we measured water with known viscosity, and simultaneously fit data for $Re<20$ to the viscous relation $\tau=\eta_l\dot\gamma$ and the data for $30 < Re < 1000$ to Eqn. 2 to obtain $Re_c=22\pm3$.  The uncertainty is one standard deviation in the statistical uncertainty on the intersection point of the two simultaneous fits.  Fitting the data for Ga and eGaIn to the same functions in the same range of $Re$ and using the known $Re_c$ gives $\eta_{\mbox{\scriptsize Ga}}=(1.92 \pm 0.08)\times 10^{-3} Pa\cdot s$ for Ga and $\eta_{\mbox{\scriptsize eGaIn}}=(1.86\pm0.09)\times10^{-3}Pa \cdot s$ for eGaIn, which are consistent with  literature values$^{1, 10}$. The uncertainty here indicates one standard deviation in the statistical uncertainty on the viscosity from the simultaneous fit of the data to the two regimes. This result demonstrates that the intrinsic liquid viscosities can be measured with higher precision by using a Reynolds scaling collapse when stress resolution limits of rheometers make low-Re measurements difficult.

This critical Reynolds number $Re_c=22\pm3$ where inertial forces begin to contribute to energy dissipation during shear is consistent with the critical Reynolds number ${Re_c}^\prime\sim 21\pm 2$ where the reduction rate of the oxide layer in acid bath increased (Sec. III. A. 1).  This is not surprising, since as flows transition from laminar to non-laminar with increasing $Re$ due to increasing inertial forces, vortices form which enhance transport and mixing of fluid. This mixing can expose more liquid metal to the surface to enhance oxidation or reduction rates at the same time.

\subsection{Surface Tension}

 \begin{figure}[h]
\begin{center}
\includegraphics[width=82mm]{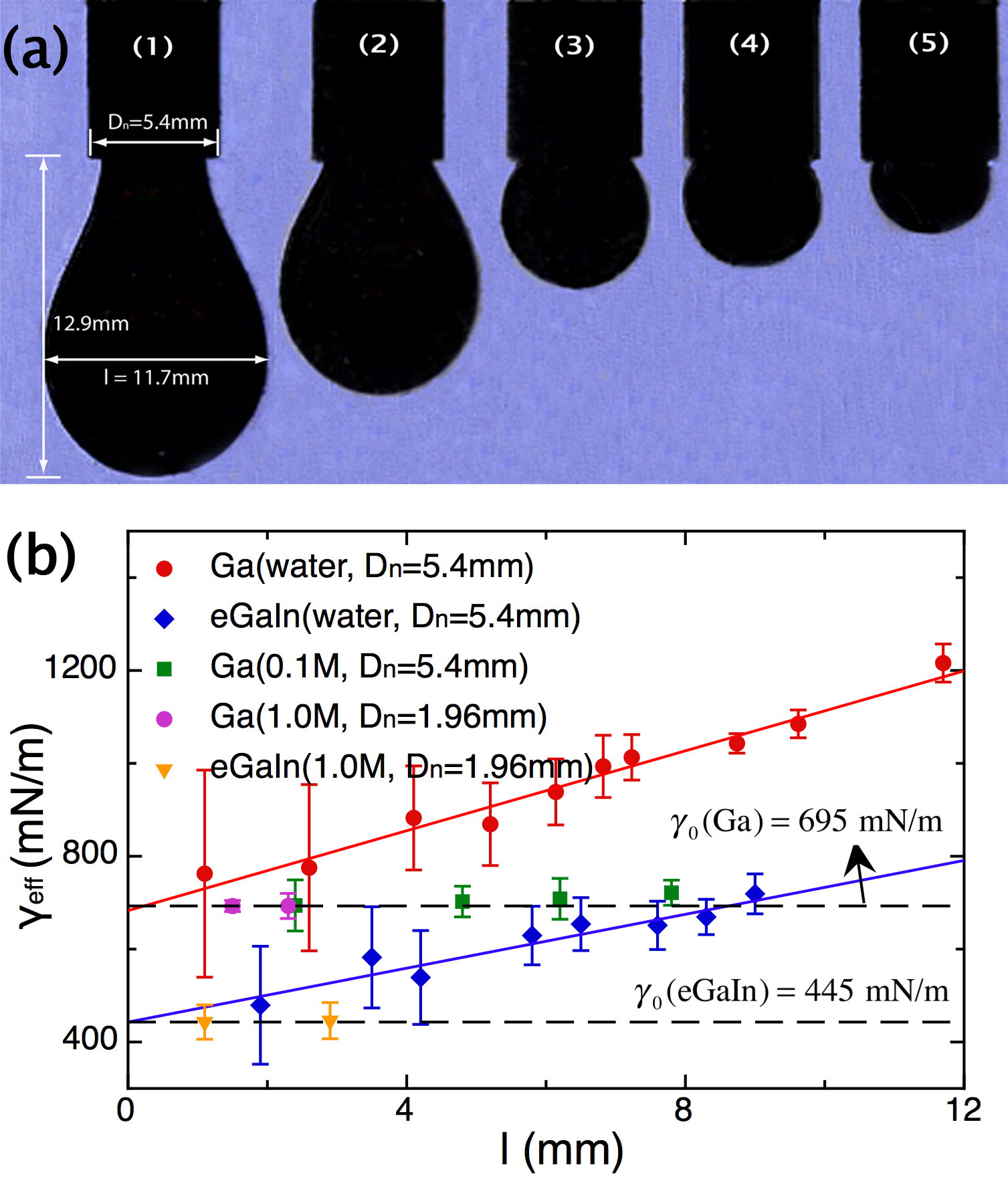}
\end{center}
\caption{Effective surface tension measurements of Ga and eGaIn using the pendant drop method. (a). Pendant droplets of liquid Ga submerged in water. (b). Measured $\gamma_{\mbox{\scriptsize eff}}$ vs drop diameter $l$ for Ga and eGaIn in different HCl baths concentrations shown in the key.  Solid lines:  fits of $\gamma_{eff} = \tau_0 l/4+\gamma_0$, which give the pure liquid surface tension $\gamma_0$ (dashed lines) and the contribution from the oxide layer $\tau_0$, which is found to be equivalent to the yield stress $\tau_y$ from Fig.~4.}
\end{figure}

In this section, we show that the oxide layer on a liquid metal can affect the effective surface tension in addition to the effective viscosity.  We performed pendant drop experiments, a standard technique for measuring surface tension.  Droplets of liquid metal were formed on the end of a nozzle of diameter $D_n=5.4$ mm and 1.96 mm.  Backlit images of Ga droplets of different sizes submerged in water are shown in Fig.~6 (a).  The effective surface tension $\gamma_{eff}$ can be obtained by fitting the equilibrium shape of the liquid metal droplet with the Laplace equation:

\begin{equation}
\Delta P_0+\Delta\rho g z=\gamma_{eff} (\frac{1}{R_1}+\frac{1}{R_2}).
\end{equation}
Here, $R_1$ and $R_2$ are the two principle radii of curvature. $\Delta \rho$ is the density difference of the two phases. The left-handed side of Eqn.~3 represents the pressure difference across the local interface, where $\Delta P_0$ is the pressure difference at a selected reference plane and $\Delta \rho g z$ stands for the hydrodynamic pressure. Here, we fit Eqn.~3 numerically using the Axisymmetric Drop Shape Analysis (ADSA) technique $^{14-16}$ to obtain the effective surface tension $\gamma_{eff}$ of Ga and eGaIn. A complete quantitative description of the ADSA method is given in the Appendix.  We plot $\gamma_{eff}$ in Fig.~6 (b) for Ga and eGaIn with different drop diameters $l$ and different HCl bath concentrations.  For the water bath ($c_{HCl}=0$), the fit value of $\gamma_{\mbox{\scriptsize eff}}$ increases with $l$, which differs from a pure liquid where $\gamma_{eff}$ should be independent of $l$.  In contrast, in the 1.0 M acid bath, the values of $\gamma_{eff}$ shown in Fig.~6 (b) are seen to be independent of $l$.  This suggests that the increase in $\gamma_{eff}$ with drop size in water is due to the oxide layer.  We  also note that in the 1.0 M HCl bath, no stable drops were formed with $l>8$ mm.  In a pure liquid the largest drop size is limited by gravitational forces overcoming surface tension to pull the drop apart.   Thus the oxide layer also allows larger droplets to remain stable.

To understand the role of the oxide skin in the pendant drop measurements, we propose that the oxide skin provides an additional stress $\tau_0$ along with the stress from surface tension to hold the drop together.  For a simple approximation, we write a spatially averaged Laplace equation (Eqn.~3) with the effective surface tension split into pure liquid surface tension and yield stress terms 

\begin{equation}
  \Delta P=\frac{2 \gamma_{\mbox{\scriptsize eff}}}{R}\simeq\frac{2 \gamma_{\mbox{\scriptsize 0}}}{R}+\tau_0,
  \end{equation}

\noindent where $\gamma_0$ is the surface tension of the pure liquid and $R$ is the average radius of curvature of the droplet.  Estimating this as $R\approx l/2$ yields an expression for the effective surface tension as a function of drop size

\begin{equation}
\gamma_{\mbox{\scriptsize eff}}\simeq \frac{\tau_0 l}{4}+\gamma_0 \ .
\end{equation}

\noindent  To test this size-dependence, we fit Eqn.~5 to the data in Fig.~6 (b), varying both $\tau_0$ and $\gamma_0$ as fit parameters. For Ga in water, the slope gives $\tau_0=172\pm12 \mbox{Pa}$, which is consistent with the yield stress measured in the rheometer without an acid bath $\tau_y= 189 \pm 12 \mbox{Pa}$ (Fig.~4).  Similarly, for eGaIn we obtain $\tau_0= 119\pm13 \mbox Pa$ while $\tau_y= 108 \pm 9\mbox Pa$.  These agreements indicate that the yield stress $\tau_y$ of the oxidized liquid metals is responsible for the larger, size-dependent effective surface tension in the pendant drop measurements.  In the limit of $l=0$ or large $c_{HCl}$, the effective surface tension in Eqn.~5 reduces to the surface tension of the pure liquid.  Fits of Eqn.~3 yield $\gamma_0(\mbox{Ga})=695\pm23 \mbox {mN/m}$ and $\gamma_0(\mbox{eGaIn})=445 \pm 19 \mbox {mN/m}$, which are in a excellent agreement with the pendant drop measurements in the acid bath shown in Fig.~6(b), as well as with literature values (Ga: $715\mbox{mN/m}$; eGaIn: $444\mbox{mN/m}$ )$^{2,3,10}$.  These quantitative agreements confirm that we can rewrite Eqn.~5 as $\gamma_{\mbox{\scriptsize eff}}=\tau_y l/4+\gamma_0$, where the effective surface tension in mechanical measurements is the linear sum of contributions from the liquid surface tension and the yield stress.

We note that since the yield stress due to the oxide layer is a surface effect, it is not size-independent like an intrinsic property, but rather it should scale as $\tau_y = k/(l/2)=2k/l$ where $k$ is the effective stiffness of the oxide layer.$^{10,11}$ This would change Eq.~5 to $\gamma_{eff} \simeq k/2+\gamma_0$. In our experiments, the gap size in the rheological test and the pendant drop size were similar, so it was not necessary to correct for the size dependence of the yield stress. 

\subsection{Wetting}

\begin{figure}[h]
\begin{center}
\includegraphics[width=72mm]{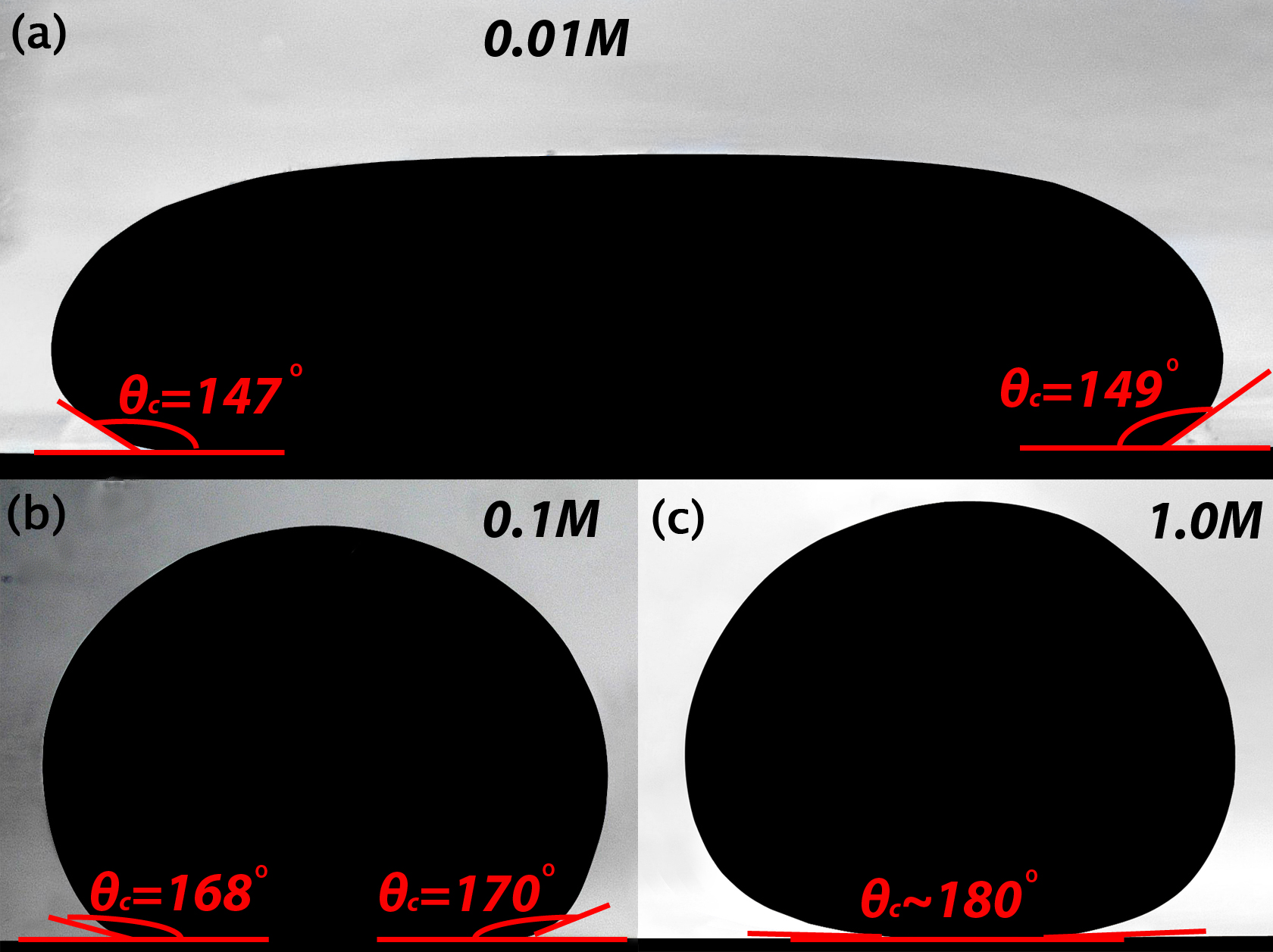}
\end{center}
\caption{Measurement of the contact angle $\theta_c$ for droplets of Ga on a glass surface submerged at different HCl concentrations labeled in the images. 
}
\end{figure}

We quantify the wetting properties of Ga and eGaIn on a glass microscope slide using the contact angle $\theta_c$. The pre-oxidized Ga and eGaIn droplets were placed on the slide with a pipette, then allowed to equilibrate with the surrounding acid for 30 min.  To avoid variations in the hydrostatic pressure from experiment to experiment, we used fixed volumes of liquid metal (250 $\mu$l) and acid (500 ml) in each experiment.  In each case, the droplets were pressed down against the glass surface using a spoon.  For $c_{HCl} > 0.2M$, the drops spontaneously bounced back to a spherical shape after removing the external force. However, at lower HCl concentrations --where the oxide skin also contributes significantly to the yield stress and pendant droplet measurements -- the droplet does not bounce back and the contact angle can vary significantly in repeated measurements.  Figure 7 shows images of these Ga droplets on glass surfaces after this procedure in acid baths with different $c_{\mbox{\scriptsize HCl}}$.  
\begin{figure*}[t]
\centering
\includegraphics[width=125mm]{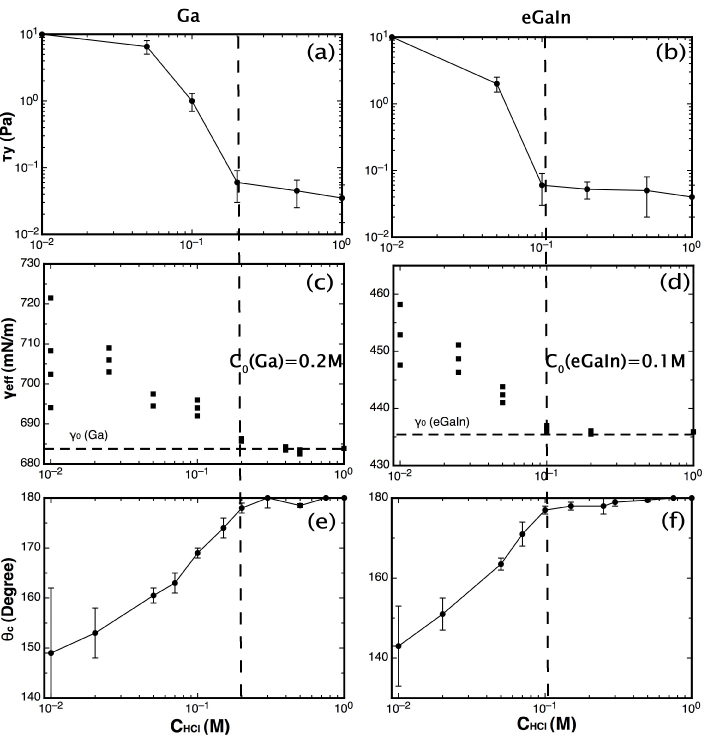}
\caption{Summary of physical parameters as a function of HCl concentration $c_{\mbox\scriptsize HCl}$ for Ga (a,c,e) and eGaIn (b,d,f). (a,b) Yield stress $\tau_y$.  (c,d) effective surface tension $\gamma_{eff}$.  Different points correspond to different pendant drop sizes, where upper points generally correspond to larger pendant drops.  (e,f) contact angle $\theta_c$.  Vertical dashed lines:  critical HCl concentration $c_{HCl} = 0.2$M for Ga and $0.1$M for eGaIn which separates pure liquid behavior from oxide skin effects.}
\end{figure*}
We measure the contact angle between the droplet and glass surfaces on both sides of the image several times and take the average as $\theta_c$. For each concentrations, and the standard deviation gives the uncertainty. These values of $\theta_c$ are plotted in Fig.~8 (e) and (f) for Ga and eGaIn, respectively, in acid baths of different concentrations. The contact angle is seen to increase with HCl concentration up to a critical concentration where $\theta_c$ reaches about 180$^{\circ}$; this is the maximum possible contact angle, corresponding to a perfectly non-wetting droplet sitting on top of the glass surface.  This suggest that pure liquid gallium is perfectly non-wettting on glass, and only wets the glass at lower acid concentrations where some oxide layer remains.

We can compare the critical HCl concentration found for the wetting on a glass surface to the effective viscosity and surface tension measurements.  The yield stress $\tau_y$ from rheometer measurements (Fig.~4) is shown vs $c_{HCl}$ in Fig.~8 (a) and (b) for Ga and eGaIn, respectively.  Similarly, the effective surface tension $\gamma_{eff}$ is shown vs $c_{HCl}$ in Fig.~8 (c) and (d) for Ga and eGaIn, respectively.   In each case, the scaling behavior of the physical parameter ($\tau_y$, $\gamma_{eff}$, or $\theta_c$), changes at a critical HCl concentration of $c_{HCl}=0.2$ M for Ga and $c_{HCl}=0.1$ M for eGaIn.  Above this critical concentration, the material behaves as a pure liquid metal with no yield stress. The values for viscosity and surface tension are in agreement with literature values, and the liquids are perfectly non-wetting on glass.  Below the critical HCl concentration, a yield stress appears, the effective surface tension appears higher, and the materials are able to partially wet glass.  The fact that these mechanical properties are all controlled by the same critical acid concentration confirms that they are related in oxidized liquid metals.  In particular, this quantitatively shows that the wetting ability of Ga and eGaIn is dependent on oxidation.

\section{Summary and Conclusion}

Our results demonstrate that the mechanical properties of Ga and eGaIn can be finely and reversibly controlled by adjusting the level of oxidation with an acid bath. Oxidation of liquid metals exposed to air leads to an increase in the shear stress (Fig. 2) and also produces a large yield stress (Fig. 4). This yield stress can be eliminated by surrounding the metal with acid, which prevents further oxidation from contact with air and reduces existing oxidation from the surface (Fig. 3). High shear rates enhance the oxidation and reduction process (Figs. 2, 3). We showed that these chemical reactions accelerate once a critical Reynolds number is exceeded that signals the onset of enhanced mixing due to vortical flows (Fig. 5).

The limited stress resolution of typical rheometers makes it difficult to resolve the viscosity of low-viscosity fluids in the laminar flow regime. To determine the intrinsic viscosity of liquid metals from rheometry data we introduced a new technique (Fig. 5). It takes advantage of the fact that, for fixed measuring geometry, all Newtonian fluids must exhibit the same scaling relation between dimensionless stress and Reynolds number. Therefore, a calibration with a fluid of known viscosity is sufficient to determine the viscosity of other Newtonian fluids in the same experimental set-up from data at higher Reynolds number.

Pendant drop measurements showed that an oxide layer increases the surface tension compared to that of the pure liquid (Fig. 6). The resulting effective surface tension can be expressed as
$\gamma_{\mbox{\scriptsize eff}} \approx \tau_y l/4+\gamma_0$, a linear combination of the pure liquid surface tension $\gamma_0$ and a contribution from the same yield
stress $\tau_y$ from rheometry (Fig. 4). This oxide-dependent increase in effective surface tension may explain the significant variation in values of the surface tension for Ga reported in the literature$^{11, 12}$.
  
To characterize wetting, we measured the contact angle of liquid metals on a glass surface, and found that the oxide layer enhances wetting  by reducing the contact angle. Without the oxide layer, liquid Ga and eGaIn are perfectly non-wetting (contact angle 180$^{\circ}$) on a glass surface (Fig. 7). Remarkably, for effective surface tension, effective viscosity, and contact angle, the scaling behavior changed at the same critical concentration of the acid bath (Fig. 8). Below this critical acid concentration the oxide layer resulted in a yield stress, large effective surface tension, and improved wetting, while above the critical concentration the intrinsic liquid properties were recovered. This provides a quantitative connection between oxidation and wetting properties of liquid metals. It also demonstrates a relationship between effective surface tension and viscosity that is unusual for Newtonian fluids.
\section{Acknowledgement}

We thank Justin Burton for providing the \emph{Mathematica} script for the ADSA method. This work is supported by the NSF MRSEC program under DMR-0820054.

\vspace{0.2in}

\appendix
\section*{{ Appendix: Axisymmetric Drop Shape Analysis (ADSA)}}

Here we summarize how to apply the ADSA method, as described in Ref. 17, to our sample.  The geometry in our experiment is an axisymmetric pendant droplet, which is described by the apex coordinate system displayed in Fig.~9 (a). 
\begin{figure*}[t]
\begin{center}
\includegraphics[width=145mm]{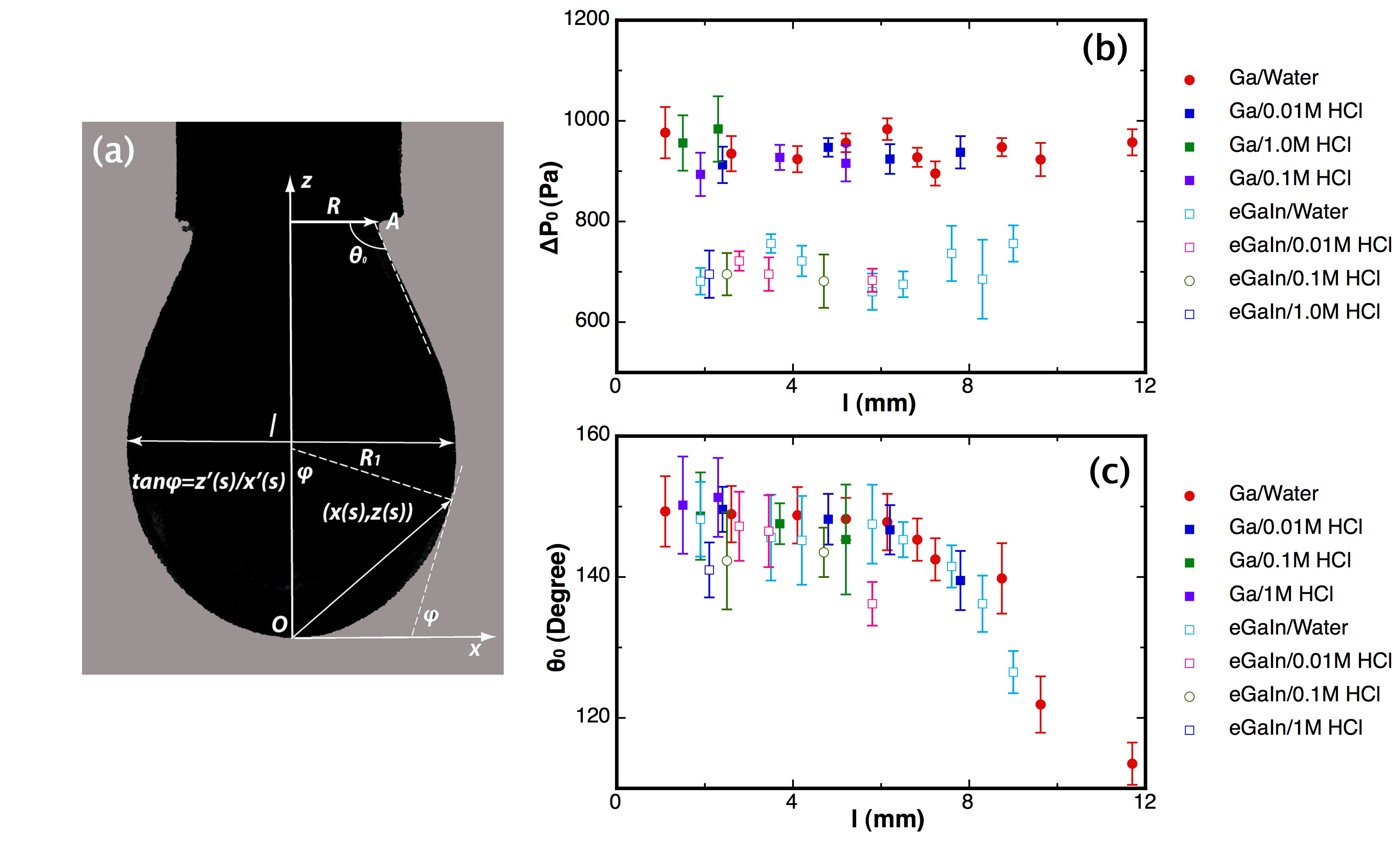}
\end{center}
\caption{\textbf{ADSA method}. (a). The apex coordinate system used in ADSA. Here, the coordinates $(x,z)$ are expressed as parametric form in term of $s$. Therefore, the local slope of $(x(s),z(s))$ is given by $\tan \varphi=x^\prime(s)/y^\prime(s)$. (b). $\Delta P_0$ vs. cross diameter $l$ for Ga (eGaIn) in the solution with different acid concentrations. (c). $\theta_0$ vs. cross diameter $l$ for Ga (eGaIn) in the solution with different acid concentrations.}
\end{figure*}
Mathematically, the shape of the axisymmetric drop can be presented by a parameter $s$ with constant differential arc length $dl=\sqrt{[x^\prime (s)^2+y^\prime (s)^2]}\:ds=Lds\:\:\:(0\leq s \leq 1)$. The arc length between points $O$ and $A$ is then $L=\int_{0}^{1}\sqrt{[x^\prime (s)^2+y^\prime (s)^2]}\:ds$. Using the curvature formula for parametric coordinates, the expression for $R_1$ and $R_2$ becomes

\begin{equation}
R_1=\frac{x^\prime(s) \sqrt{x^\prime(s)^2+y^\prime(s)^2}}{z^\prime(s)},\:\:\:\:\:\:\:\:\:\:\:\:\:\:\:\:\:\:\:\:\:\:\:\:R_2=\frac{(x^\prime(s)^2+y^\prime(s)^2)^{3/2}}{x^\prime(s) y^{\prime\prime}(s)-y^\prime(s)x^{\prime\prime}(s)}, 
\end{equation}
where the sign of $R_2$ is determined by the normal direction. For numerical consideration, the physical parameters in Laplace's Equation (Eqn.~3) can be scaled by nozzle radius $R$, surface tension $\gamma$ and density difference $\Delta\rho$. As a consequence, the Eq. 3 can be rewritten equivalently as coupled differential equations
\begin{align}
\widetilde x^{\prime\prime}(s)&=\widetilde z^{\prime\prime}(s)(2\widetilde L\widetilde P-\widetilde LB_oz(s)+\frac{\widetilde {z}^\prime(s)}{\widetilde {r}(s)}),\\
\widetilde z^{\prime\prime}(s)&=-\widetilde x^{\prime\prime}(s)(2\widetilde L\widetilde P-\widetilde LB_oz(s)+\frac{\widetilde {z}^\prime(s)}{\widetilde {r}(s)}),
\end{align}
 where $\widetilde x=x/R$, $\widetilde z=z/R$, $\widetilde L=L/R$, $\widetilde P=\Delta P_0 R/(2 \gamma)$ and $B_o=\Delta \rho gR^2/\gamma$ is the gravitational Bond number. In order to compute Eqs.~7 and 8 numerically, the boundary condition for contact angle $\theta_0$ and pressure difference $\Delta p$ at the edge of nozzle (point A in Fig.~9 (a)) need to be determined. Further, we utilized a \emph{Mathematica} code of the ADSA method to search for the best $\theta_0$, $\Delta P_0$ and surface tension $\gamma$, so that the fitted drop shape has the least mean square error. To test the validity of this method, we applied it to pendant droplets of water, where we obtained $\gamma_{water}=71.3 \pm 2.3 mN/m$ (literature value: 72.8 mN/m).

Another benchmark can be carried out by checking the value of fitted boundary condition in different measurements. Fig.~9 (b) \& (c) shows the size-dependence of $\Delta P_0$ and $\theta_0$, respectively. For both Ga and eGaIn, $\Delta P_0$ collapse onto constant values for different acid concentrations: $\Delta P_0 (\mbox{Ga}) \sim 900\mbox{Pa}$ and $\Delta P_0 (\mbox{eGaIn}) \sim 700 \mbox{Pa}$. These results are consistent with the hydrostatic pressure in the syringe with liquid height $\approx 13$ mm.  On the other hand, $\theta_0$ does depend on $l$, especially when the drop size becomes large ($l>6mm$).  This transition may be determined by the capillary length where gravitational stresses overcome surface tension.  Here, it is important that $\theta_0$ is not the same as $\theta_c$ measured in Sec. III. C.  At the contact point, the liquid metal is actually touching the nozzle edge instead of a solid plane. Therefore, the liquid is pinned around the edge and can freely vary with the shape of the droplet.

\end{spacing}
\end{document}